\documentclass[pdftex,twocolumn,epjc3]{svjour3}          
\RequirePackage{flushend}
\RequirePackage[numbers,sort&compress]{natbib}
\RequirePackage[colorlinks,citecolor=blue,urlcolor=blue,linkcolor=blue]{hyperref}
\usepackage{amsmath,lipsum,amssymb}
\usepackage{cuted}
\usepackage{latexsym}
\usepackage{amsfonts}
\usepackage{graphicx}
\usepackage{mathrsfs}
\usepackage{epstopdf}

\usepackage{orcidlink}

\journalname{Eur. Phys. J. C}
\begin{document}
\title{Inflationary Models with Gauss–Bonnet Coupling in Light of ACT Observations}
\author{Yigan Zhu \orcidlink{0009-0000-8707-3306} \thanksref{e1,addr1} \and  Qing Gao\orcidlink{0000-0003-3797-4370}\thanksref{e2,addr2} \and Yungui Gong\orcidlink{0000-0001-5065-2259}\thanksref{e3,addr1}  \and Zhu Yi\orcidlink{0000-0001-7770-9542}\thanksref{e4,addr4}}
\thankstext{e1}{e-mail: 2311690065@nbu.edu.cn}
\thankstext{e2}{e-mail: gaoqing1024@swu.edu.cn}
\thankstext{e3}{e-mail: gongyungui@nbu.edu.cn}
\thankstext{e4}{e-mail: yz@bnu.edu.cn   (corresponding author)}

\institute{Institute of Fundamental Physics and Quantum Technology, Department of Physics, School of Physical Science and Technology, Ningbo University, 818 Fenghua Rd, Ningbo, Zhejiang 315211, China \label{addr1} \and School of Physical Science and Technology, Southwest University, 2 Tiansheng Rd, Chongqing 400715, China  \label{addr2} \and Faculty of Arts and Sciences, Beijing Normal University, 18 Jinfeng Rd,  Zhuhai 519087, China \label{addr4}}
\date{Received: date / Accepted: date}

\maketitle
\begin{abstract}
Recent analyses combining Atacama Cosmology Telescope (ACT) data with other cosmological datasets report a higher scalar spectral index $n_s$, creating tension with a wide range of inflationary models.
Since a Gauss–Bonnet term with a coupling function $\xi(\phi) = 3\lambda/[4V(\phi)]$ leaves $n_s$ nearly unchanged (up to a field rescaling) while reducing the tensor-to-scalar ratio $r$ by a factor $(1-\lambda)$,
so choosing $(1-\lambda)$ sufficiently small effectively removes $r$ as a limiting observable, making it easier for inflationary models to satisfy the latest observational constraints and alleviating this tension.
Applying this mechanism to chaotic inflation, E-models, T-models, and hilltop inflation, we find that broad regions of parameter space become consistent with the latest ACT-based cosmic microwave background (CMB) constraints.
These results  demonstrate that Gauss–Bonnet couplings can help bring a broad class of inflationary models into  agreement with current CMB measurements.

\end{abstract}

\section{Introduction}
Inflation has become a cornerstone of modern cosmology, providing a compelling mechanism to solve the flatness, horizon, and monopole problems of the standard Big Bang model, 
while simultaneously generating the primordial  perturbations that seed large-scale structures and leave imprints as anisotropies in the cosmic microwave background (CMB) \cite{Starobinsky:1980te, Guth:1980zm, Linde:1981mu, Albrecht:1982wi}. 
These primordial  perturbations are commonly characterized by two key observables: the scalar spectral index $n_s$, which quantifies the scale dependence of scalar perturbations, 
and the tensor-to-scalar ratio $r$, which measures the relative amplitude of primordial gravitational waves. 
For a given inflationary potential, both observables are typically expressed in terms of the number of $e$-folds $N$ between horizon exit and the end of inflation,  
allowing precise theoretical predictions to be directly compared with observations.
A well-known class is the so-called universal attractors \cite{Kallosh:2013tua}, 
for which the scalar spectral index takes the form $n_s = 1 - 2/N$.  
This prediction arises in many models, including E- and T-models \cite{Kallosh:2013hoa, Kallosh:2013maa}, $R^2$ inflation \cite{Starobinsky:1980te}, and Higgs inflation with strong nonminimal coupling \cite{Kaiser:1994vs, Bezrukov:2007ep}. 
For $N = 60$, the universal attractor yields $n_s = 0.9667$, 
which lies in excellent agreement with the {\it Planck} 2018 result   $n_s = 0.9649 \pm 0.0042$ \cite{Planck:2018jri}.

However, recent observations from the Atacama Cosmology Telescope (ACT) \cite{ACT:2025fju, ACT:2025tim}, when combined with other datasets , report a higher value of $n_s$ compared to {\it Planck} alone. A joint analysis of ACT and {\it Planck} data (denoted as P-ACT) yields $n_s = 0.9709 \pm 0.0038$, while including CMB lensing and Dark Energy Spectroscopic Instrument (DESI) baryon acoustic oscillation (BAO) measurements \cite{DESI:2024uvr, DESI:2024mwx} (denoted as P-ACT-LB) further increases this to $n_s = 0.9743 \pm 0.0034$ \cite{ACT:2025fju, ACT:2025tim}. These latest observational data disfavor the universal attractors at approximately the $2\sigma$ level, thereby creating significant tension for a broad class of inflationary models   that predict them. 
Several approaches have been proposed to address the tension between the latest observational data and inflationary models. These include relaxing the strong-coupling limit in nonminimally coupled inflationary models of the form $\xi f(\phi) R$ \cite{Kallosh:2025rni, Gao:2025onc}, incorporating reheating dynamics \cite{Liu:2025qca, Haque:2025uis, Zharov:2025evb, Haque:2025uri, Drees:2025ngb, Ballardini:2024ado,Ye:2025idn,SidikRisdianto:2025qvk,Zharov:2025zjg,Chakraborty:2025oyj}, and investigating alternative inflationary frameworks \cite{Kim:2025dyi,Wolf:2025ecy,He:2025bli, Gialamas:2025kef, Frob:2025sfq, Dioguardi:2025vci, Brahma:2025dio, Berera:2025vsu, Aoki:2025wld, Dioguardi:2025mpp, Salvio:2025izr,  Gialamas:2025ofz,  Peng:2025bws, Pallis:2025nrv,Katsoulas:2025mcu,Byrnes:2025kit,Maity:2025czp,Addazi:2025qra,Mondal:2025kur,Odintsov:2025wai,Yogesh:2025wak, Zahoor:2025nuq,Mohammadi:2025gbu,Odintsov:2025eiv,Oikonomou:2025xms,Hai:2025wvs,Kuralkar:2025zxr}. For a recent overview, see Ref. \cite{Kallosh:2025ijd}. 

In this work, we propose an alternative resolution based on a Gauss–Bonnet term  coupled  to the inflation field via $\xi(\phi) = 3\lambda / [4 V(\phi)]$ \cite{Guo:2009uk,Jiang:2013gza}. This coupling leaves the scalar spectral index $n_s$ identical to that of the corresponding canonical model with the same potential under a rescaled field, while suppressing the tensor-to-scalar ratio $r$ by a factor of $(1 - \lambda)$. 
As a result, $r$ can  be made sufficiently small   without significantly affecting $n_s$, effectively removing the tensor-to-scalar ratio as a constraint in model selection. 
The comparison with observational data can then focus solely on $n_s$, substantially relaxing observational tensions and enabling a wide range of inflationary models to remain compatible with the recent P-ACT-LB measurements. 
For additional recent work on the Gauss–Bonnet couplings, see e.g.  Refs. \cite{ Yi:2018gse,Odintsov:2025kyw,Odintsov:2023aaw,Odintsov:2020mkz, Oikonomou:2024etl,Nojiri:2017ncd, Gangopadhyay:2022vgh, Khan:2022odn, Yogesh:2025hll,Yogesh:2024zwi,Yogesh:2024mpa,Luo:2024vls,Hussain:2025vbo}.

We apply this mechanism to several representative   inflationary models, including chaotic inflation, E- and T-models, and hilltop inflation. For each model, we compute the predictions for $n_s$ and $r$, determine the allowed parameter space, and compare with the latest observational constraints.  Our results indicate that the Gauss–Bonnet coupling can help alleviate the tension between inflationary models and current CMB constraints. Related studies on reconciling inflationary potentials with the P-ACT-LB data using a Gauss–Bonnet term 
involving  specific choices for  the coupling function,  the potential,  or both   can be found in Ref. \cite{Odintsov:2025wai}, which also incorporates the GW170817 constraints, and in Refs.\cite{Yogesh:2025wak, Zahoor:2025nuq}, where the coupling function is taken to have a hyperbolic or exponential form.

The remainder of this paper is organized as follows. In Sec. \ref{sec:GB}, we review the inflationary dynamics with a Gauss–Bonnet term and derive the expressions for the perturbation spectra. In Sec. \ref{sec:model}, we
apply the formalism to specific inflationary potentials, and compare them with the observational data.  Our conclusions are summarized in Sec. \ref{sec:con}.

\section{Inflation with  Gauss-Bonnet  coupling}
\label{sec:GB}
\subsection{The background}

The action of the inflation model with a Gauss-Bonnet coupling is given by  
\begin{equation}
    S=\frac{1}{2}\int \sqrt{-g}d^{4}x\bigg[R-g^{\mu\nu
    }\partial_\mu\phi\partial_\nu\phi-2V(\phi) -\xi(\phi)R^{2}_{GB}\bigg],    
\end{equation}
where $R^{2}_{GB}=R_{\mu\nu\rho\sigma}R^{\mu\nu\rho\sigma}-4R_{\mu\nu}R^{\mu\nu}+R^{2}$ is the Gauss-Bonnet term,  $\xi(\phi)$ is the Gauss-Bonnet coupling function,  $V(\phi)$ denotes the potential.  We adopt natural units with $c = \hbar = 1/(8\pi G) = 1$. 

For a spatially flat Friedmann–Robertson–Walker (FRW) universe,  the background equations of motion are
\begin{gather}
    \label{T00}
    6H^2=\dot{\phi}^2+2V+24\dot{\xi}H^3,\\
    \label{Tij}
    2\dot{H}=-\dot{\phi}^2+4\ddot{\xi}H^2+4\dot{\xi}H(2\dot{H}-H^2),\\
    \label{phi}
\ddot{\phi}+3H\dot{\phi}+V_{,\phi}+12\xi_{,\phi}H^2(\dot{H}+H^2)=0,
\end{gather}
where $H=\dot a /a$ is the Hubble parameter with a(t) being the cosmic scale factor. A dot denotes a derivative with respect to cosmic time $t$,  e.g., $\dot{\phi}=d\phi/dt$, and a subscript comma represents an ordinary derivative with respect to the scalar field, e.g., $V_{,\phi} \equiv dV/d\phi$.   

During slow-roll inflation, both the inflaton field $\phi$ and the coupling function $\xi(\phi)$ are assumed to vary slowly.  The slow-roll conditions are
\begin{equation} \label{sl_con}
\dot{\phi}\ll V(\phi), \; 
   |\ddot{\phi}|\ll 3 H |\dot{\phi}|, \;  4H|\dot{\xi}| \ll 1,  \; 
    |\ddot{\xi}|\ll H|\dot{\xi}|.
\end{equation}
To quantify these conditions, we define the Hubble flow parameters $\epsilon_i$ and the coupling function flow parameters $\delta_i$ as  \cite{Schwarz:2001vv, Guo:2010jr}
\begin{gather}\label{para:sr1}
    \epsilon_{1}=-\frac{\dot{H}}{H^{2}}, \quad \epsilon_{i+1}=\frac{d \ln |\epsilon_{i}|}{d  \ln a}, \quad i\geq1,\\
    \label{para:sr2}
    \delta_{1}=4\dot{\xi}H,\quad \delta_{i+1}=\frac{d \ln |\delta_{i}|}{d  \ln  a}, \quad i\geq1.
\end{gather}
Using these slow-roll parameters,  the slow-roll conditions \eqref{sl_con} can be expressed as  
\begin{equation}
    \epsilon_1\ll1, \quad |\epsilon_2| \ll1, \quad |\delta_1|\ll1, \quad |\delta_2|\ll1.
\end{equation}
In this regime, the background equations  \eqref{T00}- \eqref{phi} simplify to  
\begin{gather}
    H^2\approx \frac{1}{3}V,\label{slowroll1}\\
    \dot{H}\approx-\frac{1}{2}\dot{\phi}^2-2\dot{\xi}H^3,\label{slowroll2}\\
    \dot{\phi}\approx-\frac{1}{3H}(V_{,\phi}+12\xi_{,\phi}H^4).\label{slowroll3}
\end{gather}
Using these slow-roll background equations, the slow-roll parameters can be rewritten as
\begin{gather}\label{eps:V}
    \epsilon_1\approx\frac{Q}{2}\frac{V_{,\phi}}{V}, \quad   \epsilon_2\approx-Q\left(\frac{V_{,\phi\phi}}{V_{,\phi}}-\frac{V_{,\phi}}{V}+\frac{Q_{,\phi}}{Q}\right),\\
    \label{delta:V}
    \delta_{1}\approx-\frac{4}{3}\xi_{,\phi}QV, \quad 
    \delta_{2}\approx-Q\left(\frac{\xi_{,\phi\phi}}{\xi_{,\phi}}+\frac{V_{,\phi}}{V}+\frac{Q_{,\phi}}{Q}\right),
\end{gather}
where we have defined 
\begin{equation}\label{def:Q}
    Q=\frac{V_{,\phi}}{V}+\frac{4}{3}\xi_{,\phi}V.
\end{equation}
The number of $e$-folds $N$ from horizon exit to the end of inflation is given by
\begin{equation}\label{efold:phi}
    N= -\int_{t_e}^t H dt \approx \int_{\phi_e}^\phi \frac{d\phi}{Q},
\end{equation}
where the subscript $e$ denotes the value at the end of inflation.

\subsection{The perturbation}
The Mukhanov-Sasaki equation for the scalar perturbation is \cite{Mukhanov:1985rz,Sasaki:1986hm,Hwang:1999gf,Cartier:2001is,Hwang:2005hb,DeFelice:2011zh}
\begin{equation}
    v_k'' +\left(c_s^2 k^2-\frac{z_s''}{z_s} \right)v_k=0,
\end{equation}
where a prime denotes the derivative with respect to the conformal time $\tau = \int a^{-1} dt$.   The effective sound speed $c_s$ and the function $z_s$ are given by
\begin{gather}
    c_s^2=1-\Delta^2\frac{2\epsilon_1+\frac{1}{2}\delta_1(1-5\epsilon_1-\delta_2)}{F},\\
    z_s^2=a^2\frac{F}{(1-\frac{1}{2}\Delta)^2},
\end{gather}
where the auxiliary parameters $\Delta$ and $F$ are defined as 
\begin{align}
 \Delta &= \delta_1/(1-\delta_1),\\
 F &=2\epsilon_1-\delta_1(1+\epsilon_1-\delta_2)+3\Delta \delta_1/2. 
\end{align}
The effective mass term $z_s''/z_s$ can be written as
\begin{equation}
    \frac{z_s''}{z_s} =\frac{1}{\tau^2}\left(\nu-\frac{1}{4}\right),
\end{equation}
with the parameter $\nu$ given by
\begin{align}
    \nu=\frac{3}{2}+\epsilon_1+\frac{2\epsilon_1\epsilon_2-\delta_1\delta_2}{4\epsilon_1-2\delta_1}.    
\end{align}  
Assuming the Bunch–Davies vacuum and evaluating the solution at the  horizon crossing $c_s k = aH$, the power spectrum for the scalar perturbation is obtained as 
\begin{equation}
    \begin{aligned}
    \mathcal{P_R}&=\frac{k^3}{2\pi^2}\left|\frac{v_k}{z}\right|^2 \\
    &=2^{2\nu-3}\left[\frac{\Gamma(\nu)}{\Gamma(3/2)}\right]^2 (1-\epsilon_1)^{2\nu-1}\\\
    & \left.\times \frac{(1-\Delta/2)^2}{Fc_s^3}\left(\frac{H}{2\pi}\right)^2 \left(\frac{c_s k}{aH}\right)^{3-2\nu}\right|_{c_s k=aH},
\end{aligned}
\end{equation}
where $\Gamma(x)$ denotes the Gamma function.  The scale spectral index  is \cite{Guo:2010jr}
\begin{align}\label{ns:equation}
    n_s -1 =\frac{d \ln \mathcal{P_R}}{d \ln k}=-2\epsilon_1 -\frac{2\epsilon_1\epsilon_2-\delta_1\delta_2}{2\epsilon_1-\delta_1}.
\end{align}

Similarly, for tensor perturbations, the Mukhanov-Sasaki equation is \cite{Hwang:1999gf,Cartier:2001is,Hwang:2005hb,DeFelice:2011zh} 
\begin{equation}
    \frac{d^2 u^{b}_k}{d\tau^2} +\left(c_T^2 k^2-\frac{z_T''}{z_T} \right)u^{b}_k=0,
\end{equation}
where ``b" stands for the ``+" or ``$\times$" polarizations and 
\begin{equation}
   z_T^2= a^2(1-\delta_1), \quad  c_T^2 =1 +\Delta(1-\epsilon_1-\delta_2).
\end{equation}
The power spectrum for the tensor perturbation is given by
\begin{equation}
  \begin{aligned}
    \mathcal{P_T}&=\frac{k^3}{2\pi^2}\sum_{b=+,\times}\left|\frac{2u_k^b}{z_T}\right|^2\\
    &=\frac{2^{2\mu}}{(1-\delta_1)c_T^3}\left[\frac{\Gamma(\mu)}{\Gamma(3/2)}\right]^2 \\
    &\times \left. (1-\epsilon_1)^{2\mu-1} \left(\frac{H}{2\pi}\right)^2  \left(\frac{c_T k}{aH}\right)^{3-2\mu} \right|_{c_T k=aH},
\end{aligned}  
\end{equation}
with $\mu=3/2+\epsilon_1$.   The tensor-to-scalar ratio is \cite{Guo:2010jr}
\begin{align}\label{r:equation}
    r=\frac{\mathcal{P_T}}{\mathcal{P_R}}=16\epsilon_1-8\delta_1,
\end{align}
and the  tensor spectral index is  
\begin{equation}
 n_T=\frac{d \ln\mathcal{P_T}}{d \ln k}=-2\epsilon_1.
\end{equation}

\section{The model}
\label{sec:model}
In general situations, the slow-roll parameters $\epsilon_i$ and $\delta_i$ are independent. In this paper, we set the condition 
\begin{equation}\label{del:eps1}
    \delta_1 = 2\lambda \epsilon_1,
\end{equation}
with $0<\lambda<1$. 
Substituting this condition into the definitions of the slow-roll parameters \eqref{para:sr1} and \eqref{para:sr2}, we have
\begin{equation}\label{del:pes2}
\delta_{i+1} = \epsilon_{i+1}, \quad i\geq 1.
\end{equation}
Using Eqs. \eqref{del:eps1} and \eqref{del:pes2}, the scalar spectral index \eqref{ns:equation} and tensor-to-scalar ratio \eqref{r:equation} become 
\begin{align}\label{ns:equation2}
    n_s -1 =-2\epsilon_1 -\epsilon_2,\\
    r= 16(1-\lambda)\epsilon_1.
\end{align}
Thus the scalar spectral index coincides with that of the canonical case, while the tensor-to-scalar ratio $r$ is suppressed by a factor $1-\lambda$. In other words, under the relation \eqref{del:eps1}, the Gauss–Bonnet term suppresses $r$ without affecting $n_s$, in terms of the Hubble flow slow-roll parameters.   

Using the relations \eqref{eps:V} and \eqref{delta:V}, the condition \eqref{del:eps1} is equivalent to \cite{Yi:2018gse}
\begin{equation}\label{cor:relation}
    \xi(\phi) = \frac{3\lambda}{4V(\phi)}.
\end{equation}

Higher order curvature corrections, in particular the Gauss–Bonnet coupling, naturally arise in string theory. 
In the low-energy effective action of the heterotic   string theory, 
the Gauss–Bonnet term typically couples to the dilaton field with an exponential function \cite{Callan:1985ia,Gross:1986mw,Gasperini:1996fu}.
For an exponential potential, the string-motivated Gauss–Bonnet coupling takes the form given in Eq. \eqref{cor:relation}, which leads to power-law inflation \cite{Guo:2009uk,Jiang:2013gza}.
Furthermore, the coupling \eqref{cor:relation} implies the condition \eqref{del:eps1}, resulting the reduction to the tensor-to-scalar ratio.

Since the Hubble slow-roll parameters are not directly expressed in terms of the potential, in the following, we rewrite the results using the potential slow-roll parameters. In terms of the standard potential slow-roll parameters in the canonical inflation model, 
\begin{equation}
    \epsilon_V =\frac{1}{2}\left(\frac{V_{,\phi}}{V}\right)^2~,~\eta_V=\frac{V_{,\phi\phi}}{V},
\end{equation}
the scalar spectral index and the tensor-to-scalar ratio become
\begin{gather}\label{gb:ns}
  n_s-1= (1-\lambda) \left(2\eta_V-6\epsilon_V\right), \\
  \label{gb:r}
  r= 16(1-\lambda)^2\epsilon_V,
\end{gather}
and the expression for $Q$ in  Eq.~\eqref{def:Q} reduces to
\begin{equation}
        Q=(1-\lambda)\frac{V_{,\phi}}{V}.
\end{equation}

To the first order of the slow-roll parameters,  the  relation of $e$-folding number $N$ given by Eq. \eqref{efold:phi} becomes
\begin{equation}\label{efold:phi2}
N=\frac{1}{1-\lambda} \int_{\phi_e}^\phi \frac{V}{V_{,\phi}}d\phi.
\end{equation}

To further simplify the analysis, we introduce a field redefinition through
\begin{equation}
    \phi =\sqrt{1-\lambda} \varphi, 
\end{equation}
and define the new potential
\begin{equation}
   U(\varphi)= V(\phi)=V[\phi(\varphi)].
\end{equation}
In terms of the new field $\varphi$ and potential $U(\varphi)$, to the first order of the slow-roll parameters, the scalar spectral index \eqref{gb:ns} and tensor-to-scalar ratio \eqref{gb:r} reduce  to
\begin{gather}\label{gb:nsU}
  n_s-1 =  \left(2\eta_U-6\epsilon_U\right), \\
  \label{gb:rU}
  r =  16(1-\lambda) \epsilon_U,
\end{gather}
where the new slow-roll parameters  are defined as
\begin{equation}
    \epsilon_U =\frac{1}{2}\left(\frac{U_{,\varphi}}{U}\right)^2, \quad \eta_U=\frac{U_{,\varphi\varphi}}{U},
\end{equation}
and the $e$-folding number $N$ given in Eq. \eqref{efold:phi2} takes the form
\begin{equation}
    N = \int_{\varphi_e}^\varphi \frac{U}{U_{,\varphi}}d\varphi.
\end{equation}
Therefore, with the Gauss-Bonnet coupling function of the form $\xi(\phi) = 3\lambda/[4V(\phi)]$, the scalar spectral index remains identical to that from the  corresponding canonical model with the same potential and a rescaled field, while the tensor-to-scalar ratio is suppressed by a factor of $(1 - \lambda)$. As a result, $r$ can be made sufficiently small without affecting  $n_s$. This effectively removes the observational constraint on the tensor-to-scalar ratio, allowing model comparisons with data to focus solely on the scalar spectral index. In this way, one degree of observational tension is eliminated, making it significantly easier for inflationary models to comply with current data, such as the recent constraints from P-ACT-LB data \cite{ACT:2025fju, ACT:2025tim}:
\begin{equation}\label{act:constraints}
    n_s = 0.9743\pm 0.0034.
\end{equation}
It should be noted that, in the extreme limit $\lambda \to 1$, the model may suffer from fine-tuning issues. Moreover, since our results are derived under the first-order slow-roll approximation, if $1-\lambda$ becomes smaller than the first-order slow-roll parameters, the expressions for the scalar spectral index $n_s$, Eq. \eqref{gb:nsU}, and the tensor-to-scalar ratio $r$, Eq. \eqref{gb:rU}, are no longer applicable, and higher-order corrections should be taken into account. However, in many cases, the magnitude of $\lambda$ does not need to be close to 1; reducing the tensor-to-scalar ratio  by a factor of 2 to 3 is sufficient to match observations, i.e., $\lambda\sim 2/3$. In practical scenarios, this eliminates the two aforementioned issues.

\subsection{Chaotic inflation}
\begin{figure}[htbp]
    \centering
    \includegraphics[width=1.0\linewidth]{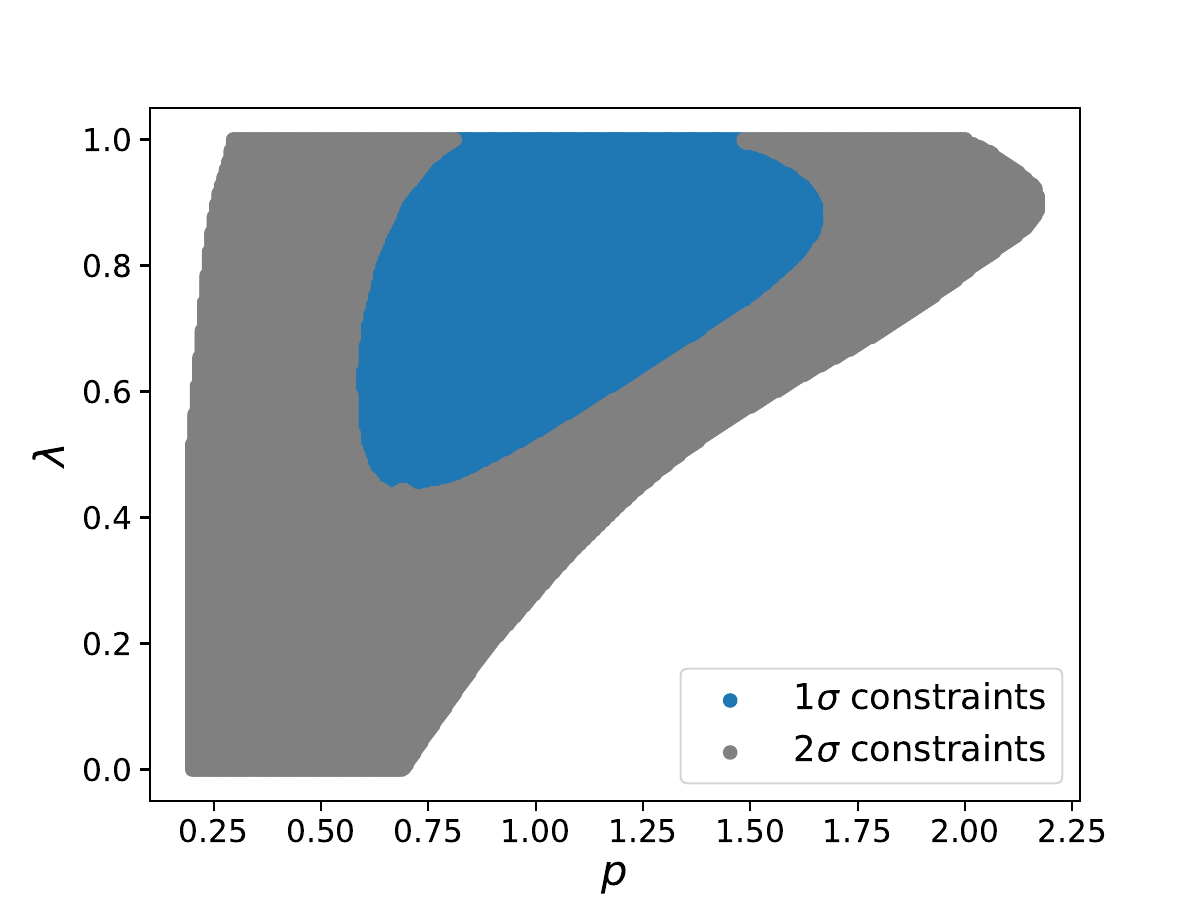}
    \caption{Constraints on chaotic inflation with a Gauss–Bonnet coupling for $N=60$. 
The blue and gray regions correspond to parameter values predicting $n_s$ and $r$ consistent with the $1\sigma$ and $2\sigma$ confidence regions from the P-ACT-LB-BK18 data, respectively. The P-ACT-LB-BK18 data are shown in Fig. \ref{fig:comparison}.}
    \label{fig:chaotic}
\end{figure}

\begin{figure}[htbp]
    \centering
    \includegraphics[width=1.0\linewidth]{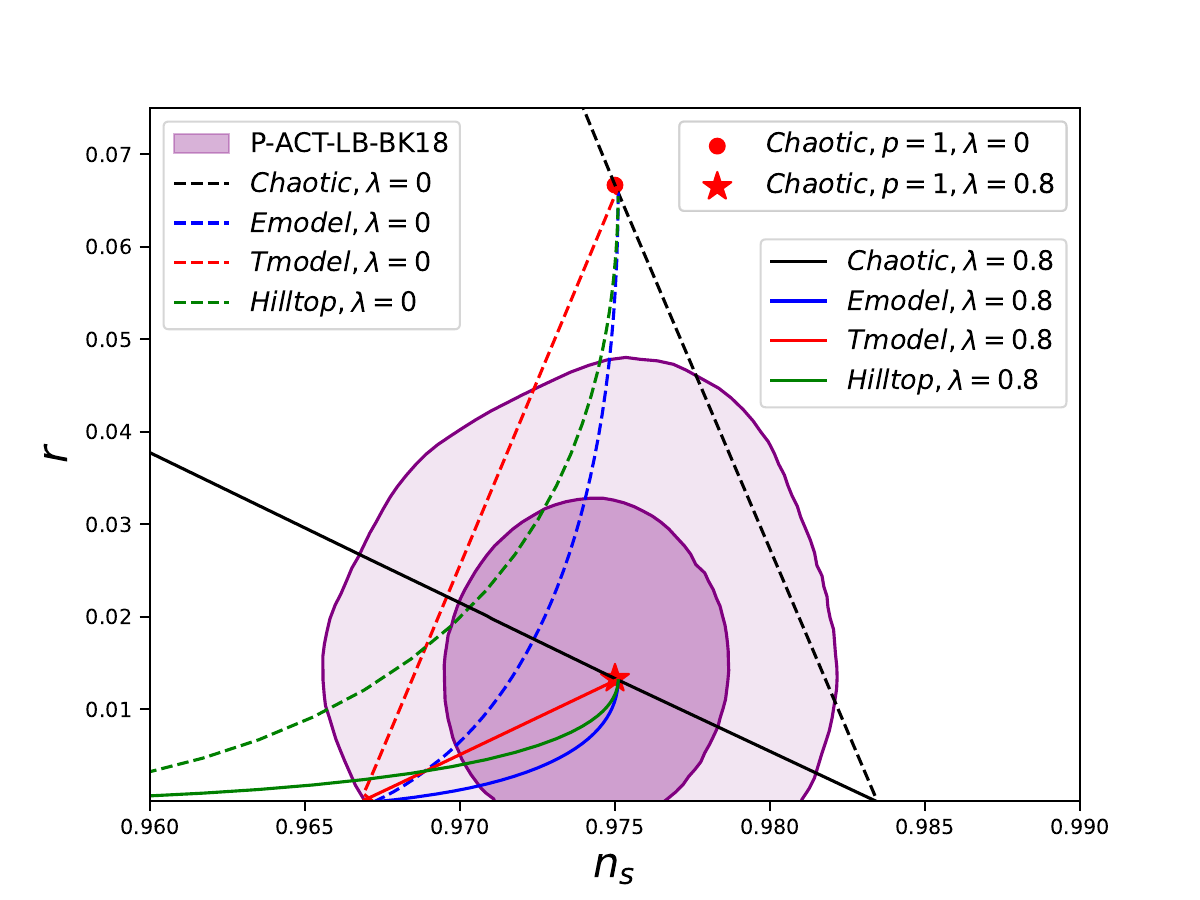}
    \caption{Comparison between the model predictions and the observational data. The purple regions indicate the $1\sigma$ and $2\sigma$ confidence regions from the P-ACT-LB-BK18 data. Dashed curves correspond to inflationary models without Gauss-Bonnet coupling ($\lambda = 0$), while solid curves represent models with Gauss-Bonnet coupling with $\lambda = 0.8$. The black, blue, red, and green curves show the predictions from the chaotic inflation model, the E-model with $n = 1/2$, the T-model with $n = 1/2$, and the hilltop inflation model with $p = 4$, respectively. The red star and red   dot indicate the predictions of the $p = 1$ chaotic inflation model with and without Gauss-Bonnet coupling, respectively.  The figure illustrates how Gauss-Bonnet coupling shifts the model predictions toward the observationally favored region.}
    \label{fig:comparison}
\end{figure}

For chaotic inflation with a monomial potential of the form \cite{Linde:1983gd}
\begin{equation}
    V(\phi)= V_0 \phi^p,
\end{equation}
with the help of the Gauss-Bonnet term, the  effective  potential is 
\begin{equation}
    U(\varphi) = U_0  \varphi^p,
\end{equation}
where $U_0 =V_0 (1-\lambda)^{p/2} $.
The predictions for the scalar spectral index $n_s$ and the tensor-to-scalar ratio $r$ are then
\begin{equation}\label{chaotic:pred}
    n_s =1-\frac{p+2}{2 (N+\tilde{n})}, \quad r= \frac{4(1-\lambda) p}{N+\tilde{n}},
\end{equation}
with $\tilde{n}=p/4$ for $2/3<p<2$, and $\tilde{n} = |p-1|/2$ for other cases.  
Taking $N=60$,  to be consistent with the P-ACT-LB constraint on $n_s$, given in Eq. \eqref{act:constraints}, the index of the chaotic inflation should satisfy 
\begin{equation}
    0.68<p<1.51.
\end{equation}
To be consistent with the joint constraints on $n_s$ and $r$ obtained from the P-ACT-LB dataset combined with the B-mode polarization measurements from the Background Imaging of Cosmic Extragalactic Polarization  (BICEP)/Keck Array experiments (BK18) \cite{BICEP:2021xfz}, denoted as P-ACT-LB-BK18, we show in Fig.~\ref{fig:chaotic} the allowed parameter space of chaotic inflation with the Gauss–Bonnet coupling for $N=60$. The parameter values in the blue and gray regions predict $n_s$ and $r$ consistent with the $1\sigma$ and $2\sigma$ confidence regions from the P-ACT-LB-BK18 data, respectively, indicated by the purple contours, as shown in Fig. \ref{fig:comparison}. Compared with using only the scalar spectral index $n_s$, the joint constraint with $r$ broadens the allowed range of the power index $p$.

The direct comparison between the model predictions for different power indices $p$, as given in Eq.~\eqref{chaotic:pred}, and the observational constraints from P-ACT-LB-BK18 is shown in Fig.~\ref{fig:comparison}.   The dashed curves represent predictions from inflationary models without Gauss-Bonnet coupling (i.e., $\lambda = 0$), while the solid curves correspond to models with Gauss-Bonnet coupling with $\lambda = 0.8$. In all cases, the number of $e$-folds is fixed to $N = 60$. The black curves denote the predictions from the chaotic inflation model. Without the Gauss-Bonnet term, chaotic inflation is disfavored by the observational data. However, with Gauss-Bonnet coupling, the tensor-to-scalar ratio is significantly reduced, bringing the model predictions into agreement with current constraints.

\subsection{$\alpha$ attractors}
For the $\alpha$-attractor E-model with a potential of the form \cite{Kallosh:2013maa,Carrasco:2015rva}
\begin{equation}
      V(\phi)=V_0 \left[1-\exp \left(-\sqrt{\frac{2}{3 \alpha }} \phi \right)\right]^{2 n}, 
\end{equation}
the presence of the Gauss-Bonnet coupling leads to an effective potential of the same form,
\begin{equation}
      U(\varphi)=V_0 \left[1-\exp \left(-\sqrt{\frac{2}{3 \tilde{\alpha} }} \varphi \right)\right]^{2 n}, 
\end{equation}
where the effective parameter $\tilde{\alpha}$ is related to the original $\alpha$ via 
\begin{equation}\label{new:alpha}
    \tilde{\alpha} = \frac{\alpha}{1-\lambda}.
\end{equation}
The resulting expressions for the scalar spectral index $n_s$ and the tensor-to-scalar ratio $r$ are \cite{Yi:2016jqr}
\begin{gather}
\label{emodnseq1}
n_s=1+\frac{8 n }{3 \tilde{\alpha} \left[ g(N,n,\tilde{\alpha})+1\right]}-\frac{8  n (n+1)}{3\tilde{\alpha}  \left[ g(N,n,\tilde{\alpha})+1\right]^2},\\
\label{emodreq1}
r=\frac{64(1-\lambda)  n^2}{3 \tilde{\alpha} \left[g(N,n,\tilde{\alpha})+1\right]^2},
\end{gather}
where  the function $g(N, n, \tilde{\alpha})$ depends on the model parameters and takes different forms in different parameter regimes. 

For the case  $n>1$ and $n/[3(2n-1)]<\tilde{\alpha}<4n^2/[3(n-1)^2]$, or $1/3<n<1$ and $\tilde{\alpha}>{4n^2}/[3(3n-1)^2]$, the $g$ function    is given by
\begin{equation}
  \label{emodg1}
  \begin{aligned}
 g(N,n,\tilde{\alpha})&=W_{-1}\bigg[-\left(\frac{2n}{\sqrt{3\tilde{\alpha} }}+1\right) \\
 &\times\exp \left(\frac{-4nN}{3\tilde{\alpha}}-\frac{2n}{\sqrt{3\tilde{\alpha}}}-1\right)\bigg],   
  \end{aligned}
\end{equation}
where $W_{-1}$ denotes the lower branch of the Lambert $W$ function. 
For the case  $n>1$ and $\tilde{\alpha}>{4n^2}/[3(n-1)^2]$, the $g$ function  becomes
\begin{equation}\label{emodg2}
\begin{aligned}
g(N,n,\tilde{\alpha})&=W_{-1}\bigg[-\left(\frac{2u}{3 \tilde{\alpha} }-\frac{2 n}{3 \tilde{\alpha} }+1\right)\\
&\times  \exp\left({-1-\frac{2 u+2 n (2N-1)}{3 \tilde{\alpha} }}\right)\bigg],
\end{aligned}
\end{equation}
with $u=\sqrt{6 \tilde{\alpha}  n^2+n^2-3 \tilde{\alpha}  n}$.  For other parameter choices, the $g$ function is given by
\begin{equation}
\label{emodg3}
\begin{aligned}
 g(N,n,\tilde{\alpha})&=W_{-1}\bigg[-\left(\frac{2 n}{3 \tilde{\alpha} }+\frac{2 v}{3 \tilde{\alpha} }+1\right)\\
 &\times  \exp\left({-1-\frac{2v+2n (2N+1)}{3 \tilde{\alpha} }}\right)\bigg],    
\end{aligned}
\end{equation}
where $v=\sqrt{n (3 \tilde{\alpha} -6 \tilde{\alpha}  n+n)}$.   

\begin{figure}[htbp]
    \centering
     \includegraphics[width=1.0\linewidth]{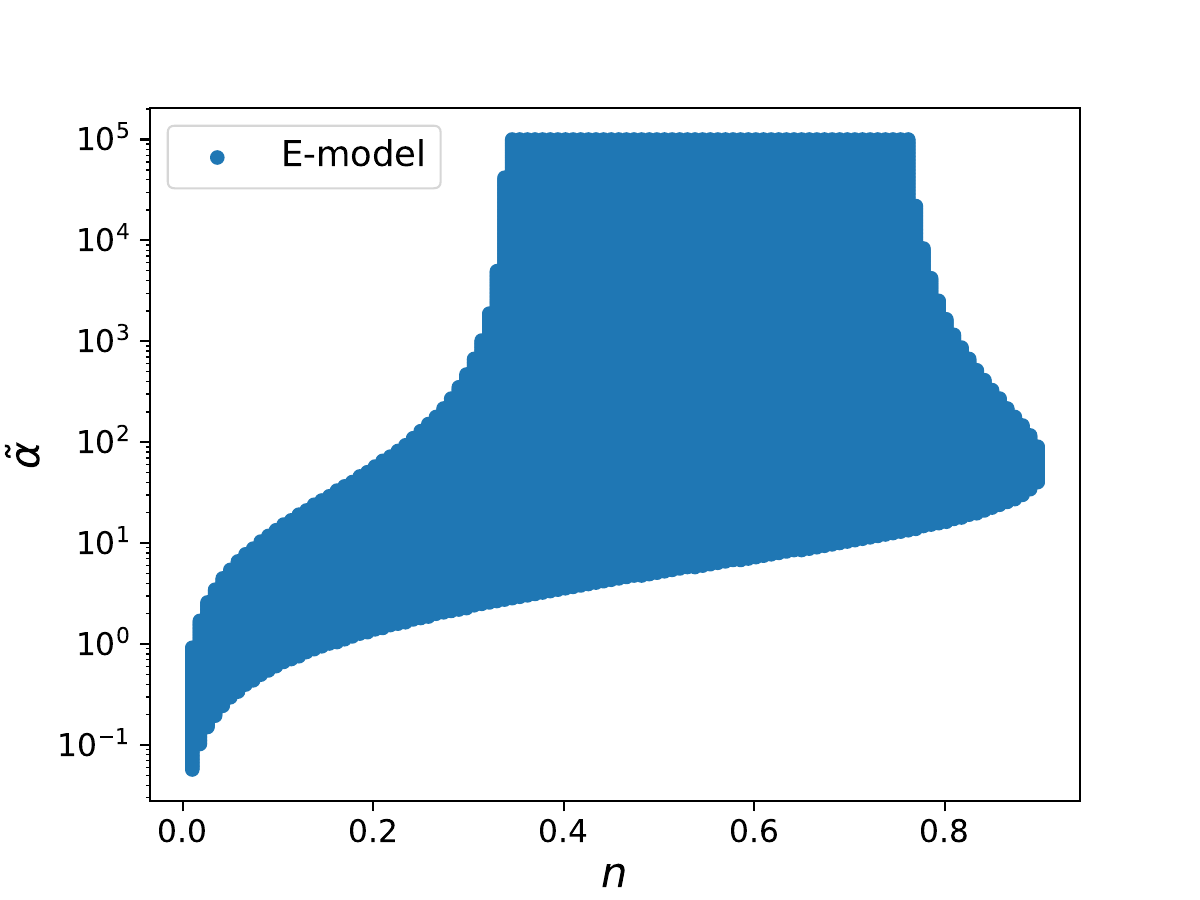}   
     \caption{Constraints on the parameters $n$ and $\tilde{\alpha}$ of the E-model, derived from the P-ACT-LB data as given in Eq.~\eqref{act:constraints}. 
    The blue region corresponds to parameter values for which the predicted scalar spectral index given by Eq.~\eqref{emodnseq1} from the E-model is consistent with the P-ACT-LB data. }
    \label{fig:Emodel}
\end{figure}

For small values of $\tilde{\alpha}$, the E-model predicts the $\alpha$-attractor form, 
\begin{equation}\label{alpha:attractor}
    n_s = 1 - \frac{2}{N}, \quad r = \frac{12(1 - \lambda)\tilde{\alpha}}{N^2},
\end{equation}
which are disfavored by the P-ACT-LB data. In the large-$\tilde{\alpha}$ limit, the E-model asymptotically approaches chaotic inflation with a power-law potential of index $p = 2n$, and the predictions reduce to those given in Eq.~\eqref{chaotic:pred}. Taking $N = 60$ and using Eqs. \eqref{act:constraints} and \eqref{emodnseq1}, the observational constraints on the parameters $\tilde{\alpha}$ and $n$ from the P-ACT-LB data are shown in Fig.~\ref{fig:Emodel}.  
For $n = 1/2$, the predictions of the E-model for different values of $\tilde{\alpha}$ are shown in Fig.~\ref{fig:comparison}, represented by the blue curves. With the inclusion of the Gauss-Bonnet term and a coupling constant $\lambda = 0.8$, the E-model becomes consistent with the P-ACT-LB-BK18 observational data for sufficiently large $\tilde{\alpha}$. 

For the $\alpha$-attractor T-model with the potential \cite{Kallosh:2013hoa, Kallosh:2013maa}
\begin{align}
    V(\phi)=V_0 \tanh ^{2 n}\left(\frac{\phi }{\sqrt{6 \alpha }}\right),
\end{align}
the effective potential in the presence of Gauss-Bonnet coupling becomes
\begin{equation}
 U(\varphi)=V_0 \tanh ^{2 n}\left(\frac{\varphi }{\sqrt{6\tilde{\alpha} }}\right),
\end{equation}
with the effective parameter $\tilde{\alpha}$   also given by Eq. \eqref{new:alpha}.  

The predictions for the scalar spectral index $n_s$ and the tensor-to-scalar ratio $r$ depend on the parameter regime. 
For the case with  $n>1$ and $(4n^2-2n\sqrt{4n^2-1})/3$ $<$ $\tilde{\alpha}<{4n^2}/[3(n^2-1)]$, $1/\sqrt{3}<n<1$ and $\tilde{\alpha}>(4n^2-2n\sqrt{4n^2-1})/3$ or  $1/3<n<1/\sqrt{3}$ and $\tilde{\alpha}>4n^2/[3(9 n^2-1)]$, 
the scalar spectral index $n_s$ and the tensor-to-scalar ratio $r$ are given by \cite{Kallosh:2013yoa,Yi:2016jqr}
\begin{gather}
\label{tmodnseq1}
   n_s=1-\frac2N
+\frac{2 N A-6 n (N-1)}{N \left[2N A +n \left(4 N^2/\tilde{\alpha}+3\right)\right]},\\  
  \label{tmodreq1}
r=\frac{48 (1-\lambda) n}{2 N A+n \left(4  N^2/\tilde{\alpha}+3\right)},  
\end{gather}
where $A=\sqrt{12 n^2/\tilde{\alpha}+9}$. 
For the case with $n>1$ and $\tilde{\alpha}>{4n^2}/[3(n^2-1)]$, the expressions become \cite{Kallosh:2013yoa,Yi:2016jqr}
\begin{gather}
\label{tmodnseq2}
n_s=1-\frac{2}{N} +\frac{8 n \left[(N-1) B  +3\tilde{\alpha} n\right]}{N \left[\left(B+ 4 n N +3\tilde{\alpha} n\right)^2-9 \tilde{\alpha} ^2\right]},  \\
\label{tmodreq2}
r=\frac{192  (1-\lambda) \tilde{\alpha}  n^2}{\left(B+ 4 n N +3\tilde{\alpha} n\right)^2-9 \tilde{\alpha} ^2}, 
\end{gather}
where $B= \sqrt{9 \tilde{\alpha} ^2+24 \tilde{\alpha}  n^2+4 n^2} -3\tilde{\alpha}n -2n $. 
For all other parameter choices, the predictions are given by \cite{Kallosh:2013yoa,Yi:2016jqr}
\begin{gather}
\label{tmodnseq3}
n_s=1-\frac2N+\frac{8 n \left[(N+1) C- 3n \tilde{\alpha}\right]}{N \left[\left(C+4 n N+3\tilde{\alpha} n\right)^2-9 \tilde{\alpha} ^2\right]},\\      
\label{tmodreq3}
r=\frac{192  (1-\lambda) \tilde{\alpha}  n^2}{\left(C+4 n N+3\tilde{\alpha} n\right)^2 -9 \tilde{\alpha} ^2},
\end{gather}
where $C=\sqrt{9 \tilde{\alpha} ^2 -24 \tilde{\alpha}  n^2 +4 n^2} -3\tilde{\alpha} n +2 n$.

\begin{figure}[htbp]
    \centering
     \includegraphics[width=1.0\linewidth]{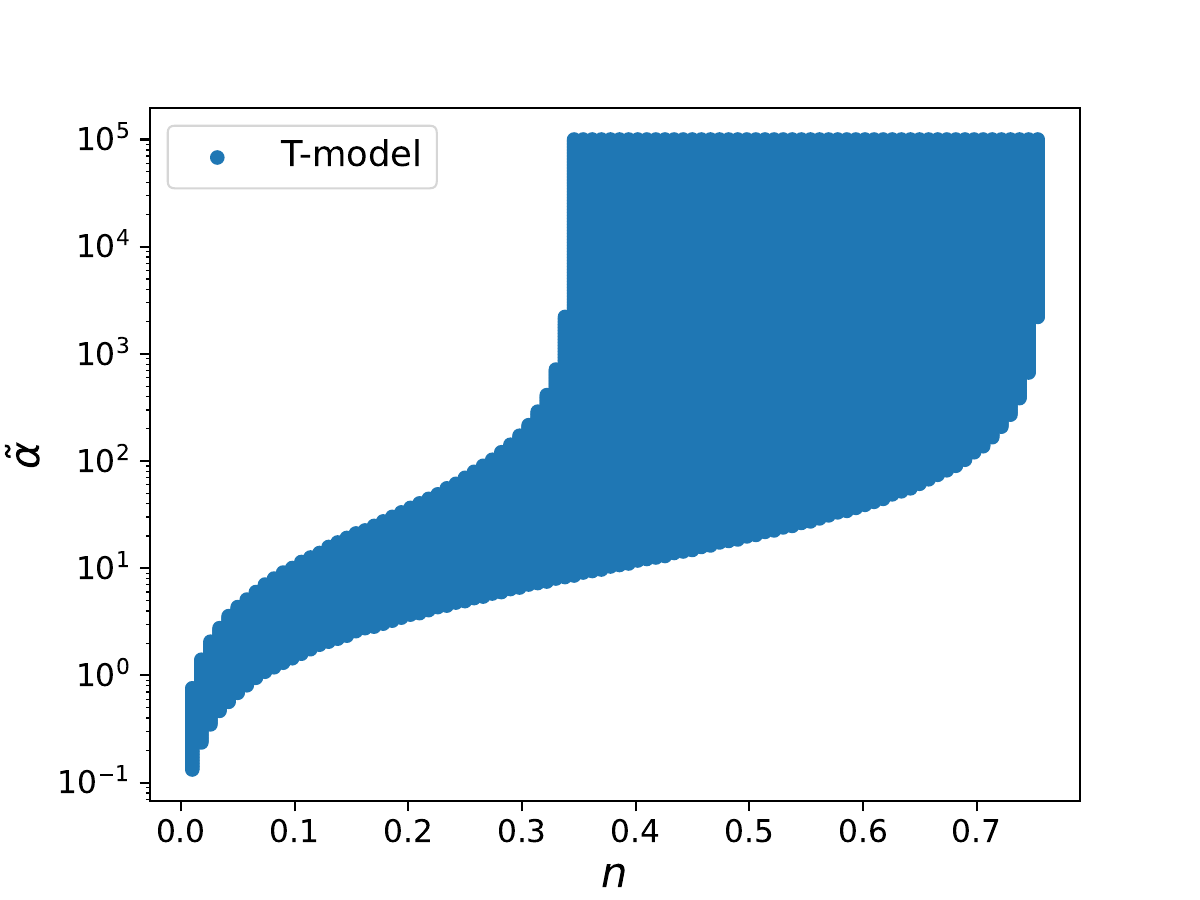}
    \caption{Constraints on the parameters $n$ and $\tilde{\alpha}$ of the T-model, derived from the P-ACT-LB data as given in Eq.~\eqref{act:constraints}.
    The blue region corresponds to parameter values for which the predicted scalar spectral index given by Eq.~\eqref{tmodnseq1} from the T-model is consistent with the P-ACT-LB data. }
    \label{fig:Tmodel}
\end{figure}

Similar to the E-model, in the small-$\tilde{\alpha}$ limit, the predictions of the T-model reduce to those of the $\alpha$-attractor form given in Eq.~\eqref{alpha:attractor}. In the large-$\tilde{\alpha}$ limit, the T-model approaches the chaotic inflation model with a power-law potential of index $p = 2n$ \cite{Kallosh:2013yoa}, and the predictions coincide with those in Eq.~\eqref{chaotic:pred}. Taking $N = 60$, the observational constraints on the parameters $\tilde{\alpha}$ and $n$ of the T-model from the P-ACT-LB data are shown in Fig.~\ref{fig:Tmodel}. For $n = 1/2$, the predictions of the T-model for different values of $\tilde{\alpha}$ are shown in Fig.~\ref{fig:comparison}, represented by the red curves. With the inclusion of the Gauss-Bonnet term and a coupling constant $\lambda = 0.8$, the T-model becomes consistent with the P-ACT-LB-BK18 observational data for sufficiently large $\tilde{\alpha}$. 

\subsection{Hilltop inflation}
For hilltop inflation with the potential \cite{Boubekeur:2005zm}
\begin{equation}
  V(\phi)=V_0 \left[1-\left(\frac{\phi }{\mu }\right)^p\right],
\end{equation}
the effective potential in the presence of Gauss-Bonnet coupling becomes
\begin{equation}
      U(\varphi)=V_0 \left[1-\left(\frac{\varphi }{\tilde{\mu} }\right)^p\right],
\end{equation}
where the effective parameter is $\tilde{\mu} = \mu / \sqrt{1 - \lambda}$. In this work, we focus on the case with $p > 2$.

The slow-roll parameters for the effective potential are given by
\begin{align}
    \epsilon_U&=\frac{p^2 \left( \varphi/\tilde{\mu} \right)^{2 p-2}}{2 \tilde{\mu} ^2 \left[1-\left(\varphi/\tilde{\mu}\right)^p\right]^2},\\
    \eta_U&=-\frac{(p-1) p \left( \varphi/\tilde{\mu}  \right)^{p-2}}{\tilde{\mu} ^2 \left[1-\left( \varphi /\tilde{\mu}\right)^p\right]}.
\end{align}
The number of $e$-folds between horizon crossing and the end of inflation is related to the field value by
\begin{equation}\label{hilltop:efold}
N = \frac{\tilde{\mu}^2}{p}\left[f(\varphi_*/\tilde{\mu})-f(\varphi_e/\tilde{\mu})\right],
\end{equation}
where 
\begin{equation}
    f(x)= \frac{1}{2}x^2 - \frac{x^{2-p}}{2-p}.
\end{equation}
Here, $\varphi_e$ and $\varphi_*$ denote the field values at the end of inflation and at horizon crossing, respectively. The value of $\varphi_e$ is determined by the condition that either $\epsilon_U(\varphi_e) = 1$ or $|\eta_U(\varphi_e)| = 1$, whichever is satisfied first.
 
The scalar spectral index and tensor-to-scalar ratio are given by
\begin{align} \label{hilltop:ns}
     n_s &=1 -\frac{3 p^2 \left( \varphi_*/\tilde{\mu} \right)^{2 p}}{\varphi_* ^2 \left[\left( \varphi_*/\tilde{\mu}  \right)^p-1\right]^2}+\frac{2 (p-1) p \left( \varphi_* /\tilde{\mu}  \right)^p}{\varphi_* ^2 \left[\left( \varphi_* /\tilde{\mu} \right)^p-1\right]},   \\
      r&=   \frac{16(1-\lambda) p^2 \left( \varphi_*/\tilde{\mu} \right)^{2 p-2}}{2 \tilde{\mu} ^2 \left[1-\left(\varphi_*/\tilde{\mu}\right)^p\right]^2}.
\end{align}
 In the general case, it is difficult to express $\varphi_*$ analytically in terms of the $e$-folding number $N$, making it hard to write $n_s$ and $r$ directly as functions of $N$.
 
However, in the small-$\tilde{\mu} \ll 1$ limit, approximate expressions can be obtained:
\begin{gather}
    n_s = 1 - \frac{2(p-1)}{(p-2)N}, \\
    r =\frac{8p^2}{\tilde{\mu}^2}\left[\frac{\tilde{\mu}^2}{p(p-2)N}\right]^{(2p-2)/(p-2)}.
\end{gather}
In this limit, the predicted scalar spectral index is smaller than the universal attractor value $n_s = 1 - 2/N$, and thus is disfavored by current observational data. Therefore, to be consistent with observations, it is necessary to go beyond the small-$\tilde{\mu}$ regime. 

In the large-$\tilde{\mu}$ limit, we have $1/\tilde{\mu}^2 \ll 1$ and $\varphi/\tilde{\mu} < 1$. The end of inflation is determined by the condition that either $\epsilon_U(\varphi_e) = 1$ or $|\eta_U(\varphi_e)| = 1$, both of which imply $\varphi_e \approx \tilde{\mu}$. 
From Eq.~\eqref{hilltop:efold}, if $\tilde{\mu}$ is sufficiently large such that $N/\tilde{\mu}^2 \ll 1$, we find $f(\varphi_*/\tilde{\mu}) \approx f(\varphi_e/\tilde{\mu})$, which leads to $\varphi_* \approx \varphi_e \approx \tilde{\mu}$. Therefore, during inflation, the field excursion is very small, and the potential can be well approximated by a linear expansion: $U(\varphi) \simeq U_0 p(1 -  \varphi/\tilde{\mu})$. This approximate potential yields  the prediction of the chaotic inflation model with $p = 1$, which  is consistent with the P-ACT-LB observational data. Therefore, in the large-$\tilde{\mu}$ limit, hilltop inflation can be compatible with current observations.

\begin{figure}[htbp]
    \centering
    \includegraphics[width=1.0\linewidth]{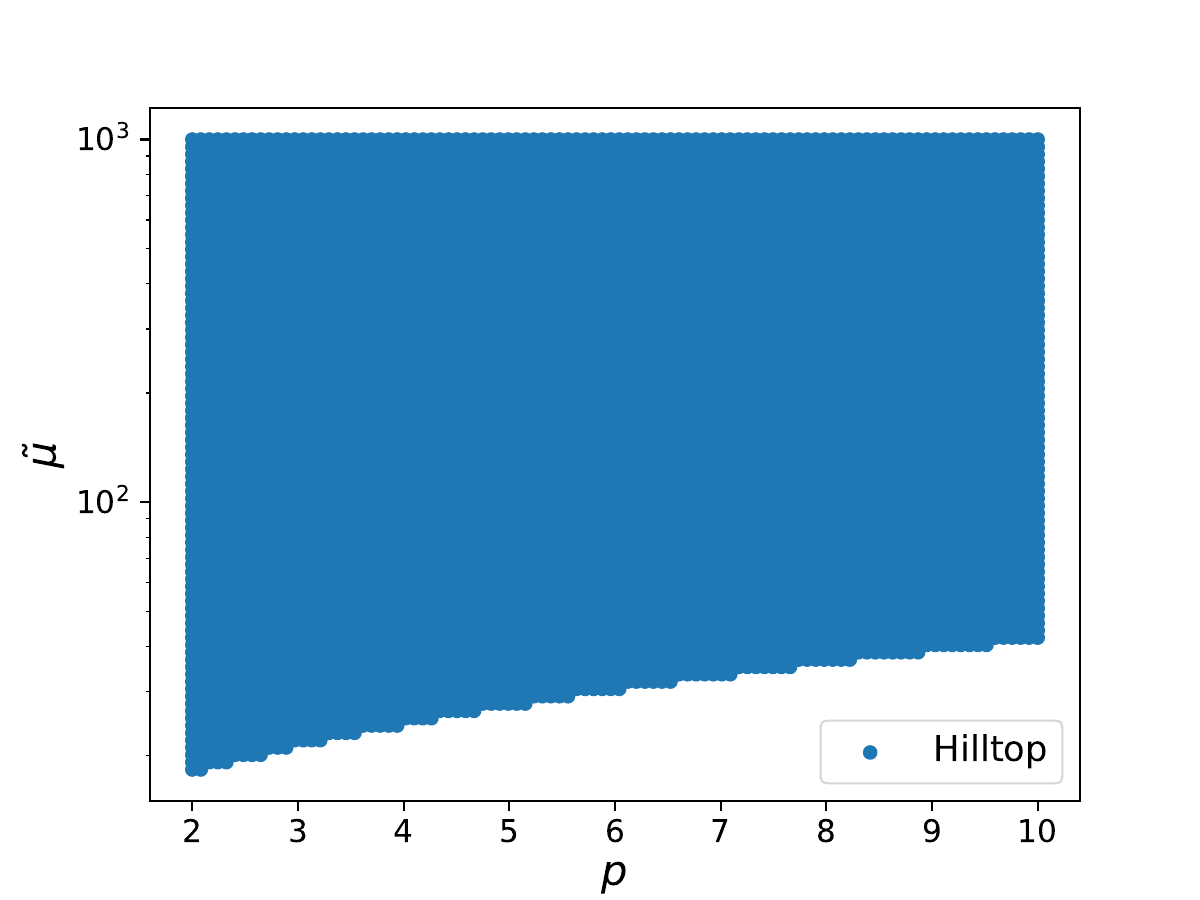}
    \caption{Constraints on the parameters $p$ and $\tilde{\mu}$ of the hilltop inflation model, derived from the P-ACT-LB data as given in Eq.~\eqref{act:constraints}. 
    The blue region corresponds to parameter values for which the predicted scalar spectral index given by Eq.~\eqref{hilltop:ns} from the hilltop inflation is consistent with the P-ACT-LB data. }
    \label{fig:hilltop}
\end{figure}

By numerically solving the scalar spectral index   for a general $\tilde{\mu}$, and comparing it with the observational constraints given in Eq.~\eqref{act:constraints}, we   obtain the constraints on the parameters of the hilltop inflation model,   shown in Fig.~\ref{fig:hilltop}.  For the case $p = 4$, the predictions of the hilltop inflation model with varying $\tilde{\mu}$ are presented in Fig.~\ref{fig:comparison}, represented by the green curves. With the inclusion of the Gauss-Bonnet term and a coupling constant $\lambda = 0.8$, the model can be brought into agreement with the P-ACT-LB-BK18 observational data for sufficiently large values of $\tilde{\mu}$.   

Similar results can also be obtained for other values in the moderate range  $0.5<\lambda<1$, which indicates that the agreement with the observational constraints does not rely on a fine-tuned choice of  $\lambda$.
\section{Discuss and Conclusion}
\label{sec:con}
Recent measurements from the Atacama Cosmology Telescope, when combined with {\it Planck}, CMB lensing, and DESI BAO data (P-ACT-LB), indicate a higher scalar spectral index, $n_s = 0.9743 \pm 0.0034$. This value is about $2\sigma$ above the {\it Planck}-only result and challenges the universal attractor prediction $n_s = 1 - 2/N$ predicted by many well-known inflationary models, including the E-model, T-model, $R^2$ inflation, and Higgs inflation with strong nonminimal coupling.

In this work, we have shown that   a Gauss–Bonnet term with a coupling of the form $\xi(\phi) =3\lambda/[4V(\phi)]$ can reconcile these models with the new data. 
This coupling function leaves the prediction for $n_s$ identical to that of the corresponding canonical model with the same potential under a field rescaling,  while suppressing the tensor-to-scalar ratio $r$ by a factor $(1 - \lambda)$. 
As a result, $r$ can be made sufficiently small, effectively removing it as a constraining observable and allowing model viability to be determined solely by $n_s$. 
We have applied this framework to several representative inflationary potentials, including chaotic inflation, E-model, T-model, and hilltop inflation, and have identified the regions of model parameter space consistent with the latest P-ACT-LB  constraints.
For chaotic inflation, agreement with the P-ACT-LB data requires $0.68 < p < 1.51$.
In the large-$\alpha$ limit of the E- and T-models, the predictions reduce to those of chaotic inflation with $p = 2n$, which is compatible with the observed $n_s$ if $0.68 < 2n < 1.51$.
In the large-$\mu$ limit of hilltop inflation, the predictions match those of chaotic inflation with $p = 1$, which are also consistent with the P-ACT-LB value of $n_s$. 
In all cases, the tensor-to-scalar ratio can be brought within observational bounds by taking $(1 - \lambda)$ sufficiently small. As an example, for $\lambda = 0.8$, chaotic inflation, the E- and T-models with $n = 1/2$ and large $\alpha$, and hilltop inflation with large $\mu$ can be made fully consistent with the P-ACT-LB-BK observational data, which incorporate the BICEP/Keck (BK18) measurements of B-mode polarization and thus include constraints on the tensor-to-scalar ratio.

Our results suggest that the Gauss–Bonnet coupling may help bring a broad class of inflationary models into better agreement with current CMB measurements. This mechanism offers a potentially useful approach for easing tensions between theoretical predictions and observational data, and its relevance is likely to extend beyond the specific examples considered here. 
We also note that, although the suppression of the tensor-to-scalar ratio becomes more effective as $\lambda \to 1$, our mechanism does not rely on extreme fine-tuning: values in the moderate range $0.5 < \lambda < 1$ are already sufficient to satisfy the current observational limits. Nevertheless, if $|1-\lambda|$ becomes smaller than the first-order slow-roll parameters, higher-order corrections should be considered beyond the approximation used in this work.

Upcoming CMB polarization experiments such as Lite (Light) satellite for the studies of B-mode polarization and Inflation from cosmic background Radiation Detection (LiteBIRD) \cite{LiteBIRD:2022cnt} and CMB-S4 \cite{CMB-S4:2016ple} will dramatically improve the sensitivity to primordial tensor modes, aiming at the level of $r \lesssim 10^{-3}$.  If primordial gravitational waves are detected and the tensor-to-scalar ratio is found in the range $10^{-2}$–$10^{-3}$, such an observation would support the Gauss–Bonnet suppression mechanism, corresponding to  $1-\lambda \sim \mathcal{O}(10^{-1})$. While this may raise concerns about fine-tuning, we note that $1-\lambda$ at this level still exceeds the first-order slow-roll parameters, so that our analytic results remain valid within the slow-roll framework.
Conversely, a null result at $r < 10^{-3}$ would push the model into a regime with $1-\lambda < \mathcal{O}(10^{-2})$, where higher-order corrections may become necessary. In addition, future CMB observations will further improve the precision of the scalar spectral index $n_s$, which may provide complementary constraints on Gauss–Bonnet inflationary scenarios.

\begin{acknowledgement}
This work is supported in part by the National Key Research and Development Program of China under Grant No. 2020YFC2201504, and the National Natural Science Foundation of China under Grant No. 12205015. 
\end{acknowledgement}


\end{document}